\documentclass[journal,onecolumn,draftcls]{IEEEtran}
\usepackage{amsmath,amsthm,mathtools}
\usepackage{graphicx}
\usepackage{color}
\usepackage{epsfig, amsmath, amsthm, amssymb,latexsym,amsmath,amsbsy, rotating}
\usepackage{amsfonts}
\usepackage{hyperref}
\usepackage{cleveref}

\usepackage{tikz}
\usetikzlibrary{matrix,positioning,fit,calc,shapes}
\definecolor{cbblue}{RGB}{159, 074, 150}

\newcommand{\Var}{\mathrm{Var}}

\DeclarePairedDelimiter\parentheses{\lparen}{\rparen}
\newcommand{\entp}[1]{\operatorname{N} \parentheses*{#1}}
\newcommand{\entpsq}[1]{\operatorname{N}^2 \parentheses*{#1}}

\DeclareFontFamily{U}{futm}{}
\DeclareFontShape{U}{futm}{m}{n}{
	<-> s * [.92] fourier-bb
}{}
\DeclareSymbolFont{Ufutm}{U}{futm}{m}{n}
\DeclareSymbolFontAlphabet{\mathbb}{Ufutm}

\DeclareMathOperator{\Tr}{trace}

\theoremstyle{plain}
\newtheorem{Theorem}{Theorem}
\newtheorem{Lemma}{Lemma}
\newtheorem{Corollary}{Corollary}

\newtheorem*{Corollary*}{Corollary}

\theoremstyle{definition}

\theoremstyle{remark}

\makeatletter
\newcounter{labelcnt}
\renewcommand{\thelabelcnt}{(\alph{labelcnt})}
\newcommand{\setlabel}[1]{%
  \refstepcounter{labelcnt}\ltx@label{lbl:#1}%
  {\text{\upshape\thelabelcnt}}%
}

\makeatother

\begin{document}

\title{Shannon Bounds for Quadratic\\ Rate-Distortion Problems}

\author{Michael~Gastpar,~\IEEEmembership{Fellow,~IEEE} and Erixhen~Sula 
\thanks{The work in this manuscript was supported in part by the Swiss National Science Foundation under Grant 200364. The work in this manuscript was partially presented at the {\it 2022 IEEE International Symposium on Information Theory}.}
\thanks{M. Gastpar is with the School of Computer and Communication Sciences, {\'E}cole Polytechnique F{\'e}d{\'e}rale de Lausanne (EPFL), Lausanne, Switzerland, e-mail: michael.gastpar@epfl.ch. Erixhen Sula was with MIT. }
}

\IEEEaftertitletext{\vspace{-1\baselineskip}\begin{center} \it Dedicated to the memory of Toby Berger (1940-2022) \end{center}}

\maketitle

\begin{abstract}
The Shannon lower bound has been the subject of several important contributions by Berger.
This paper surveys Shannon bounds on rate-distortion problems under mean-squared error distortion with a particular emphasis on Berger's techniques. Moreover, as a new result, the Gray-Wyner network is added to the canon of settings for which such bounds are known. In the Shannon bounding technique, elegant lower bounds are expressed in terms of the source entropy power.
Moreover, there is often a complementary upper bound that involves the source variance in such a way that the bounds coincide in the special case of Gaussian statistics.
Such pairs of bounds are sometimes referred to as {\it Shannon bounds.}  The present paper puts Berger's work on many aspects of this problem in the context of more recent developments, encompassing indirect and remote source coding such as the CEO problem, originally proposed by Berger,
as well as the Gray-Wyner network as a new contribution.
\end{abstract}


\IEEEpeerreviewmaketitle

\section{Introduction}

Rate-distortion theory was initiated by Shannon~\cite{Shannon:59}. Berger~\cite{BergerBook} developed the first full and still authoritative account on this important topic. 
Rate-distortion theory develops the fundamental limits of lossy data compression.
In the classic data compression problem, a single underlying source sequence needs to be compressed. From the compressed representation, a decoder produces an approximate version of the source sequence. The quality of the approximate version is assessed via a distortion measure. 
This leads to a trade-off between the rate of the compressed representation and the incurred distortion.
The corresponding fundamental limit is referred to as the rate-distortion function. The foundational result of rate-distortion theory shows that the rate-distortion function can be expressed as a constrained information minimization problem, see, e.g.,~\cite[Thm.10.2.1]{CoverThomas06}.
This is a convex optimization problem.
It can be tackled numerically~\cite{Blahut:1972,Arimoto:1972}, but it
rarely admits explicit closed-form solutions. Notable special cases for which such solutions are known include the case of Gaussian statistics when the distortion measure is the mean-squared error, and the binary case subject to Hamming distortion.
Beyond such special cases, however, one needs to either resort to numerical answers or to bounds.
The focus of the present paper concerns bounds to the rate-distortion function of a particular type.
Specifically, early on in the development of rate-distortion theory, Shannon devised an elegant way of bounding the rate-distortion function via the entropy of the source~\cite{Shannon:59}.
This argument works for all so-called {\it difference distortion measures} but has been most appreciated in the special case of quadratic distortion, that is, the mean-squared error.
Part of the reason may be that in this special case, the Shannon lower bound can often be complemented by a corresponding upper bound of exactly the same shape, but with the source variance in place of the source entropy power.
Such pairs of bounds are sometimes referred to as {\it Shannon bounds.} 
In the simplest case, discussed in~\Cref{sec:shannonbounds:classic} below, these bounds take the following shape:
\begin{align}
\frac{1}{2}\log^+{\frac{\entp{X}}{\Delta}} \leq R_X(\Delta) \leq \frac{1}{2}\log^+{\frac{\Var(X)}{\Delta}}.\label{eq:classicshannonbound:intro}
\end{align}
The main features of Shannon bounds can be summarized as follows:
\begin{enumerate}
\item They are automatically tight in the case of Gaussian statistics. 
\item Worst-case property: When fixing only source variances, Shannon bounds directly imply that Gaussian statistics are worst.
\item Best-case property: When fixing only source entropies, Shannon bounds directly imply that Gaussian statistics are best.
\item The gap between the upper and the lower bound is the Kullback-Leibler divergence between the source distribution and a Gaussian distribution with the same variance.
\end{enumerate}

\subsection{Contributions}

Berger, in his seminal book~\cite[Section 4.3]{BergerBook}, provides an excellent account of the pioneering work on Shannon bounds along with some key extensions.
In the present paper, we review Berger's many contributions to the topic and put them in the context of more recent developments.
In particular, for point-to-point problems, we encompass the conditional rate-distortion function, the Wyner-Ziv rate-distortion function, and the remote (or indirect) rate-distortion function. We show how these results are established via the techniques of Berger, most specifically, his elegant variational formula for the rate-distortion function. The latter lends itself beautifully to the development of lower bounds.

The last two sections of the paper concern multi-terminal (network) rate-distortion problems. Berger~\cite{berger-msc_lecturenotes} made foundational contributions to this important topic. A new technical contribution of our work is a pair of Shannon bounds for the so-called Gray-Wyner network given in~\Cref{thm:main}. We then end our discussion with a special case of the CEO problem. This source coding network was introduced by Berger~\cite{BergerZV:96}.

The new result,~\Cref{thm:main}, may share its apparent shape with classic results such as the one in~\Cref{eq:classicshannonbound:intro}. However, it is important to note that the techniques to establish these two lower bounds, respectively, are vastly different. The lower bound in~\Cref{eq:classicshannonbound:intro} can be established via elementary facts about entropy and a judicious maximum entropy argument. By contrast, the lower bound in our new result (\Cref{thm:main}) relies on techniques including Brascamp-Lieb-type inequalities (such as~\Cref{lemma:thm:main} below).

\subsection{Related Work}

The Shannon lower bound has been the subject of a steady trickle of work.
The main focus of this paper is on the case of {\it scalar}  sources. The Shannon lower bound has also been extended to the case of vector sources by Yamada, Tazaki, and Gray~\cite{YamadaTG:80}. Further extensions concern Markov sources, where a notable result is due to Avram and Berger~\cite{AvramB:85}, establishing the critical distortion below which the Shannon lower bound is tight.

Berger discusses the advantages of the Shannon lower bound in his book~\cite[Section 4.3.4]{BergerBook}, pointing out that the bound is interesting in the low-distortion regime.
This angle has been the subject of intense studies over the years.
Special scrutiny was reserved for the {\it asymptotic}  tightness of the Shannon lower bound, notably in the work of Linder and Zamir~\cite{LinderZ:94} and Koch~\cite{Koch:16}. A finite block-length refinement of the Shannon lower bound was developed by Kostina~\cite{Kostina:15}.

The Shannon lower bound also plays a role in a string of interesting applications. For example, Feng and Effros, in~\cite{FengE:05} and Weissman and Ordentlich, in~\cite{WeissmanO:05}, establish theorems for sources for which the Shannon lower bound is tight. Zamir, Kochman, and Erez~\cite{ZamirKE:08} leverage the Shannon lower bound in an application involving predictive coding. The Shannon lower bound also plays a pivotal role in Rose's mapping approach~\cite{Rose:94}. Nayak and Tuncel~\cite{NayakT:09} leverage Shannon lower bounds in the context of successive coding.
The Shannon lower bound has appeared in multiterminal problems, e.g., in the work of Zamir and Berger~\cite{ZamirB:99}.

Our focus in this paper is on quadratic rate-distortion problems. However, the rationale of the Shannon lower bound easily extends to difference distortion measures, which is perhaps most obvious from Berger's proof in~\cite[Section 4.3.4]{BergerBook}.
Several results are known for difference distortion measures beyond the mean-squared error.
Berger~\cite{Berger:77} pitches the Shannon lower bound (applied in the case of binary sources and Hamming distortion) against an improved bound.
Bounds akin to the Shannon lower bound can also be derived for distortion measures that are not difference distortion measures, see e.g., Linder and Zamir in~\cite{LinderZ:99} and with Zeger in~\cite{LinderZZ:99}.
Shannon lower bounds for special classes of sources are given by Koliander {\it et al.} in~\cite{KolianderPRH:16}.

\subsection{Notation}

Throughout this paper, we will use $\log(\cdot)$ to denote the natural logarithm, and for any non-negative real number $x,$ we use the shorthand $\log^+(x) = \max(0, \log(x)).$

For a continuous real-valued random variable $X,$ the variance is denoted by $\Var(X),$ the (differential) entropy by $h(X),$ and the entropy power as
\begin{align}
 \entp{X} & = \frac{e^{2h(X)}}{2 \pi e},
\end{align}
and we recall that for Gaussian random variables $X,$ we have that $\entp{X} = \Var(X).$

For two jointly continuous real-valued random variables $X$ and $W,$ we will use the following standard notation
\begin{align}
 \Var(X|W=w) & = {\mathbb E} \left[  \left.\left( X - {\mathbb E}[X|W] \right)^2 \right| W=w \right],
\end{align}
and we denote the expected value of this quantity over $W$ by the semi-standard notation
\begin{align}
  \Var(X|W) & = {\mathbb E}_{X,W} \left[  \left( X - {\mathbb E}[X|W] \right)^2 \right],
\end{align}
which is the minimum mean-squared error (MMSE) in estimating $X$ from $W.$
Moreover, we will use the notation
\begin{align}
 \entp{X|W} & = \frac{e^{2h(X|W)}}{2 \pi e},
\end{align}
where one can again see that when $X$ and $W$ are jointly Gaussian, then $\entp{X|W} =   \Var(X|W) = \sigma_X^2(1-\rho^2),$
where $\rho$ denotes the correlation coefficient between $X$ and $W.$

Moreover, we also use $\entp{X,Y}$ to denote the entropy power of the vector $(X,Y)$, that is,
\begin{align}
\entp{X,Y}=\frac{1}{2 \pi e} e^{h(X,Y)}.
\end{align}
When $X$ and $Y$ are jointly Gaussian with covariance matrix $\Sigma,$ then $\entpsq{X,Y} = \det\Sigma,$ which is the product of the eigenvalues of $\Sigma.$
To see this, recall that in this case, $h(X,Y)= \frac{1}{2}\log (2\pi e)^2 \det \Sigma.$
It is important to recall that this is {\it not}  the expected value of the squared $L_2$-norm of the random vector $(X,Y),$ which is $\Tr\Sigma,$ and thus the sum of the eigenvalues of $\Sigma.$

\section{MMSE Estimation}

A first statement akin to the Shannon lower bound can be made about the classic problem of estimation subject to the mean-squared error criterion.
Let us define the minimum mean-squared error as
\begin{align}
 \Delta_0 &= {\mathbb E} \left[ \left( X - {\mathbb E}[X|Y] \right)^2 \right].\label{eq:def:mmse}
\end{align}
Then, whenever $\Delta_0\le \entp{X},$ we have (letting $\rho$ denote the correlation coefficient between $X$ and $Y$)
\begin{align}
 \entp{X|Y} & \le \Delta_0 \le \sigma_X^2(1 - \rho^2) .\label{eqn-estimation-raw}
\end{align}
where we have equality throughout if and only if $X$ and $Y$ are jointly Gaussian.
This result can be found, e.g,, in~\cite[p.255]{CoverThomas06}.

While these bounds can be useful, we may have a preference for an even more explicit pair of bounds that we discuss next.
In this case, we specifically study the case where $Y$ is an observation of $X$ subject to {\it additive}  (independent) noise, that is,
\begin{align}
 Y &= X +Z.
\end{align}
For this case, the bounds take the following more explicit form, see e.g.~\cite[Eqn.(1.34)]{EswaranG:09}:
\begin{align}
 \frac{\entp{X}\entp{Z}}{\entp{Y}} \le \Delta_0 \le \frac{\sigma_X^2\sigma_Z^2}{\sigma_Y^2}. \label{eqn-estimation}
\end{align}
A key feature of the bounds in~\Cref{eqn-estimation-raw} and in~\Cref{eqn-estimation} is that their shape is identical;
and that variance terms in the upper bound are systematically replaced by the corresponding entropy powers to obtain the lower bound.
Therefore, in the special case of Gaussian statistics, these bounds are equal and thus tight.
We will refer to such bounds as Shannon bounds.

\section{The Quadratic Rate-Distortion Function}\label{sec:shannonbounds:classic}

In~\cite{BergerBook}, Berger makes several intricate and deep contributions to rate-distortion theory in the special case of so-called difference distortion measures.
This designates the case where the value of $d(x, y)$ only depends on the difference between $x$ and $y$ (and thus, requires $x$ and $y$ to take value in a group).
In this paper, we restrict attention to mean-squared error distortion, which is of course a special case of a difference distortion measure.
Another important example is the class of modulo-additive distortion measures over a finite group.

\subsection{The Scalar Case}

Consider the classic problem of compressing a continuous-amplitude source in a lossy fashion. Let the source be modeled by a sequence of independent and identically distributed continuous random variables, as described in detail, e.g., in~\cite{BergerBook} and~\cite[Chapter 10]{CoverThomas06}. From the compressed representation, the original source sequence needs to be reconstructed to within a prescribed distortion level $\Delta.$ How many bits of coding rate are required to attain this goal?
The answer to this question is the rate-distortion function. A careful discussion of this can be found, e.g., in~\cite{BergerBook} and~\cite{CoverThomas06}.
For the case where the distortion measure is the mean-squared error, the (so-called quadratic) rate-distortion function is given by
\begin{align}
R_{X}(\Delta):=\inf_{p(\hat{x}|x):\mathbb{E}[(X-\hat{X})^2]\leq \Delta} I(X;\hat{X}).  \label{Eq-basicRD}
\end{align}

The specific goal of the present paper is to discuss the optimization problem stated in~\Cref{Eq-basicRD}.
For most distributions $p_X(x),$ this problem cannot be solved in closed form.
It can be solved numerically, but this shall not be of particular interest here.
Instead, our interest is in {\it bounds}  on the rate-distortion function in~\Cref{Eq-basicRD}.
More specifically, we are interested in the following type of bounds, originally due to Shannon and further developed by Berger.

\begin{Theorem}\label{thm-SLB-basic}
For a continuous random variable $X$ and mean-squared error distortion $d(x,\hat{x})=(x-\hat{x})^2,$ the rate-distortion function is bounded by
\begin{align}
\frac{1}{2}\log^+{\frac{\entp{X}}{\Delta}} \leq R_X(\Delta) \leq \frac{1}{2}\log^+{\frac{\Var(X)}{\Delta}}.\label{eq:shannonbound:basic}
\end{align}
\end{Theorem}

Needless to say, this basic version of the Shannon bound is a classic exercise in many books and classes on information theory.
Its proof is usually accomplished leveraging the fact that conditioning reduces entropy together with a maximum entropy argument.
Berger, in his book~\cite[Sec.4.3]{BergerBook}, provides a rather different but equally interesting proof approach.
For completeness, we include a brief sketch of Berger's argument of the lower bound in this theorem in Appendix~\ref{app-BergerProofSLB}.

To interpret~\Cref{thm-SLB-basic}, we note that the gap between the upper and the lower bound in~\Cref{eq:shannonbound:basic} can be expressed in an instructive fashion as
\begin{align}
    \frac{1}{2}\log^+{\frac{\Var(X)}{\Delta}} - \frac{1}{2}\log^+{\frac{\entp{X}}{\Delta}} &= D(f \| g),
\end{align}
for $\Delta\le\entp{X},$
where $D(\cdot\|\cdot)$ denotes the Kullback-Leibler divergence, $f$ denotes the distribution of the random variable $X,$ and $g$ denotes a Gaussian distribution with the same variance as $X.$
This shows that the bounds in~\Cref{thm-SLB-basic} are interesting for distributions that are close to Gaussian.

\subsection{The Vector Case}

Let $(X_1,X_2)$ be distributed with full-rank covariance matrix $\Sigma.$
Suppose that this vector needs to be compressed to yield lossy descriptions at mean-squared error distortion $\Delta_1$ and $\Delta_2,$ respectively.
For this problem, the corresponding rate-distortion function is given by
\begin{align}
R_{(X_1,X_2)}(\Delta_1, \Delta_2) :=\inf_{\substack{p(\hat{x}_1,\hat{x}_2|x_1, x_2):\\\mathbb{E}[(X_1-\hat{X}_1)^2]\leq \Delta_1,\\ \mathbb{E}[(X_2-\hat{X}_2)^2]\leq \Delta_2}} I(X_1,X_2;\hat{X}_1,\hat{X}_2).  \label{Eq-RD-vector-individual}
\end{align}

For this problem, the Shannon lower bound is well known, see e.g.~\cite{ZamirB:99}, and given by the following theorem:

\begin{Theorem}\label{thm-SLB-vector-individual}
For a continuous random vector $(X_1,X_2)$ with full-rank covariance matrix $\Sigma$ and separate (coordinate-wise) mean-squared error distortion criteria, the rate-distortion function is bounded by
\begin{align}
    \min_D \frac{1}{2} \log^+ \frac{\entpsq{X_1,X_2}}{\det D}
    & \le R_{(X_1,X_2)}(\Delta_1, \Delta_2) \le \min_D \frac{1}{2} \log^+ \frac{\det \Sigma}{\det D}
\end{align}
where the minimum is over all matrices $D$ that satisfy $0 \preccurlyeq D \preccurlyeq \Sigma$ (i.e., in semi-definite ordering) and $D_{ii}\le \Delta_i.$
\end{Theorem}

As a side remark, we note that this theorem can be extended to a vector of arbitrary length.

A simple extension of the lossy source coding problem addressed by~\Cref{thm-SLB-vector-individual} is the case where a distortion constraint is only imposed on the sum, rather than individually on the components.
The corresponding rate-distortion function can be expressed simply by taking the definition in~\Cref{Eq-RD-vector-individual} and adding an outer minimization over all non-negative pairs $(\Delta_1,\Delta_2)$ whose {\it sum} is at most the desired sum distortion level.
This leads to an allocation problem known as {\it (reverse) water-filling} in the information-theoretic literature, see e.g.~\cite[Section 10.3.3]{CoverThomas06}.
\Cref{thm-SLB-vector-individual} also extends to this case by adding, both in the upper and in the lower bound, an outer optimization.

\section{The Conditional Rate-Distortion Function and the Wyner-Ziv Problem}

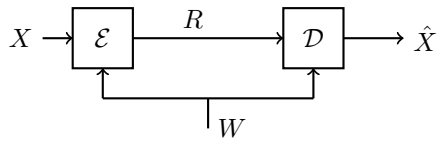
\begin{figure}[h!] 
\centering
\begin{tikzpicture}[scale=0.8]
\draw[black,thick] (0,1) rectangle (1,2);
\draw[black,thick,anchor=east] (-.5,1.5) node{$X$};
\draw[black,thick,anchor=west] (5.5,1.5) node{$\hat{X}$};
\draw[black,thick,anchor=center] (0.5,1.5) node{$\mathcal{E}$};
\draw[black,thick,anchor=south] (2,1.5) node{$R$};
\draw[->,black,thick](-0.5,1.5)--(0,1.5);
\draw[->,black,thick](1,1.5)--(3.5,1.5);
\draw[black,thick,anchor=center] (4,1.5) node{$\mathcal{D}$};
\draw[black,thick] (3.5,1) rectangle (4.5,2);
\draw[->,black,thick](4.5,1.5)--(5.5,1.5);
\draw[->,black,thick](0.5,0.5)--(0.5,1);
\draw[->,black,thick](4,0.5)--(4,1);
\draw[black,thick](0.5,0.5)--(4,0.5);
\draw[black,thick](2.25,0)--(2.25,0.5);
\draw[black,thick,anchor=west] (2.25,0) node{$W$};
\end{tikzpicture}
\caption{The conditional rate-distortion problem.} 
\label{fig:conditional}
\end{figure}

In this section, we consider two basic rate-distortion scenarios with side information, namely, the conditional rate-distortion function and the Wyner-Ziv problem.
In these problems, there is the source of interest, $X,$ and there is side information $W,$ correlated with the source.
This side information may be available both at the encoder and at the decoder, in which case it is referred to as the {\it conditional rate-distortion function.}
Or it may be available at the receiver only, in which case it is referred to as the {\it Wyner-Ziv}  rate-distortion function.
For both cases, one can give bounds of the type considered in the present paper.

The conditional rate-distortion problem is illustrated in~\Cref{fig:conditional}.
Specifically, let the source be modeled by a sequence of independent and identically distributed pairs of jointly continuous random variables, $X$ and $W,$ as described in detail in~\cite{Gray--1972}. The encoder produces a compressed representation.
From the compressed representation together with the side information sequence $W,$ the original source sequence needs to be reconstructed to within a prescribed distortion level $\Delta.$ How many bits of coding rate are required to attain this goal?
The answer to this question is the conditional rate-distortion function. A careful discussion of this can be found in~\cite{Gray--1972}.
For the case where the distortion measure is the mean-squared error, the conditional rate-distortion function is given by
\begin{align}
R_{X|W}(\Delta):=\inf_{p(\hat{x}|x,w):\mathbb{E}[(X-\hat{X})^2]\leq \Delta} I(X;\hat{X}|W).
\end{align}
Clearly, for every realization $W=w,$ this is a regular rate-distortion problem. Indeed, one can establish the following lower and upper bounds in the spirit of the bounds given in~\Cref{thm-SLB-basic}.

\begin{Theorem} \label{lem:cond-rate-distor}
The quadratic conditional rate-distortion function is bounded by
\begin{align}
\frac{1}{2}\log^+{\frac{\entp{X|W}}{\Delta}} \leq R_{X|W}(\Delta) \leq \frac{1}{2}\log^+{\frac{\Var(X|W)}{\Delta}}.
\end{align}
\end{Theorem}

\Cref{lem:cond-rate-distor} does not appear to be in the archival literature, aside from~\cite{SulaG:22isit}. A proof is included in Appendix~\ref{app:proof:lem:cond-rate-distor}. Our proof crucially leverages Berger's variational representation of the rate-distortion function as discussed in Appendix~\ref{app-BergerProofSLB}.

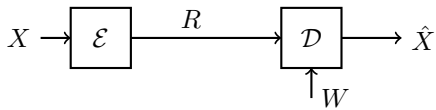
\begin{figure}[h!] 
\centering
\begin{tikzpicture}[scale=0.8]
\draw[black,thick] (0,1) rectangle (1,2);
\draw[black,thick,anchor=east] (-.5,1.5) node{$X$};
\draw[black,thick,anchor=west] (5.5,1.5) node{$\hat{X}$};
\draw[black,thick,anchor=center] (0.5,1.5) node{$\mathcal{E}$};
\draw[black,thick,anchor=south] (2,1.5) node{$R$};
\draw[->,black,thick](-0.5,1.5)--(0,1.5);
\draw[->,black,thick](1,1.5)--(3.5,1.5);
\draw[black,thick,anchor=center] (4,1.5) node{$\mathcal{D}$};
\draw[black,thick] (3.5,1) rectangle (4.5,2);
\draw[->,black,thick](4.5,1.5)--(5.5,1.5);
\draw[->,black,thick](4,0.5)--(4,1);
\draw[black,thick,anchor=west] (4,0.5) node{$W$};
\end{tikzpicture}
\caption{The Wyner-Ziv rate-distortion problem.} 
\label{fig:WZ}
\end{figure}

An interesting variant of the conditional rate-distortion function has been widely studied in the literature. In this variant, called the {\it Wyner-Ziv problem,} only the decoder has access to the side information. This is illustrated in~\Cref{fig:WZ}.
Specifically, let the source be modeled by a sequence of independent and identically distributed pairs of jointly continuous random variables, $X$ and $W,$ as described in detail~\cite{WynerZ:76}. Observing only the sequence $X,$ the encoder produces a compressed representation.
From the compressed representation together with the side information sequence $W,$ the original source sequence needs to be reconstructed to within a prescribed distortion level $\Delta.$ How many bits of coding rate are required to attain this goal?
The answer to this question is the Wyner-Ziv rate-distortion function. A careful discussion of this can be found in~\cite{WynerZ:76}.
For the case where the distortion measure is the mean-squared error, the Wyner-Ziv rate-distortion function is given by
\begin{align}
R_{X|W}^{{\mathrm{WZ}}}(\Delta):=\inf_{p(u|x), p(\hat{x}|u,w):\mathbb{E}[(X-\hat{X})^2]\leq \Delta} I(X;U|W). \label{eq:def:WZ}
\end{align}
It is immediately clear that the conditional rate-distortion function is a relaxation of this minimization problem.
Therefore, we must have $R_{X|W}(\Delta) \le R_{X|W}^{{\mathrm{WZ}}}(\Delta) .$
Hence, the lower bound given in~\Cref{lem:cond-rate-distor} remains a valid lower bound for the case of the Wyner-Ziv rate-distortion function.
For the Wyner-Ziv rate-distortion function, the following pair of bounds can be given.

\begin{Theorem} \label{lem:cond-rate-distor-WZ}
The quadratic Wyner-Ziv rate-distortion function is bounded by
\begin{align}
\frac{1}{2}\log^+{\frac{\entp{X|W}}{\Delta}} \leq R_{X|W}^{{\mathrm{WZ}}}(\Delta) \leq  \frac{1}{2}\log^+{\frac{\sigma_X^2(1-\rho^2)}{\Delta}} ,
\end{align}
where $\rho$ denotes the correlation coefficient between $X$ and $W.$
\end{Theorem}

The upper bound in this theorem is rather straightforward and follows from analyzing a judiciously chosen auxiliary $p(u|x)$ in~\Cref{eq:def:WZ}. For completeness, the full argument is outlined in Appendix~\ref{app:proof:lem:cond-rate-distor}.

It is tempting conjecture that the lower bound in~\Cref{lem:cond-rate-distor-WZ} could be improved, since the problem of~\Cref{fig:WZ} is more difficult than that of~\Cref{fig:conditional}.
But for the particular type of bounds studied in the present paper, this is not possible.
In fact, for the special case where $X$ and $W$ are jointly Gaussian, all bounds given in~\Cref{lem:cond-rate-distor} and~\Cref{lem:cond-rate-distor-WZ}
coincide. Hence, this is an alternative perspective on and proof of the well-known fact that in the case of jointly Gaussian $(X,Y),$ there is no penalty for not knowing the side information at the encoder.

Comparing the upper bounds in~\Cref{lem:cond-rate-distor} and~\Cref{lem:cond-rate-distor-WZ}, we observe that~\Cref{lem:cond-rate-distor} has the minimum mean-squared error while~\Cref{lem:cond-rate-distor-WZ} has the {\it linear} minimum mean-squared error. It is left as an open problem whether the upper bound of~\Cref{lem:cond-rate-distor} also applies in the context of~\Cref{lem:cond-rate-distor-WZ}.

\section{Remote (Indirect) Source Coding}\label{Sec-remote} 

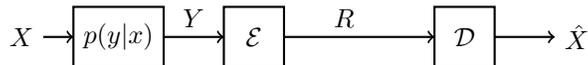
\begin{figure}[h!] 
\centering
\begin{tikzpicture}[scale=0.8]
\draw[black,thick] (-2.5,1) rectangle (-1,2);
\draw[black,thick] (0,1) rectangle (1,2);
\draw[black,thick,anchor=east] (-3,1.5) node{$X$};
\draw[black,thick,anchor=south] (-0.5,1.5) node{$Y$};
\draw[black,thick,anchor=west] (5.5,1.5) node{$\hat{X}$};
\draw[black,thick,anchor=center] (-1.75,1.5) node{$p(y|x)$};
\draw[black,thick,anchor=center] (0.5,1.5) node{$\mathcal{E}$};
\draw[black,thick,anchor=south] (2,1.5) node{$R$};
\draw[->,black,thick](-3,1.5)--(-2.5,1.5);
\draw[->,black,thick](-1,1.5)--(0,1.5);
\draw[->,black,thick](1,1.5)--(3.5,1.5);
\draw[black,thick,anchor=center] (4,1.5) node{$\mathcal{D}$};
\draw[black,thick] (3.5,1) rectangle (4.5,2);
\draw[->,black,thick](4.5,1.5)--(5.5,1.5);
\end{tikzpicture}
\caption{The Remote Rate-Distortion Problem} 
\label{fig:remote}
\end{figure}

The problem of remote source coding was originally introduced in by Dobrushin and Tsybakov~\cite{DobrushinT:62}.
It is the subject of an in-depth discussion in Berger's book on rate-distortion theory~\cite[Section 3.5]{BergerBook}.
The setting is illustrated in~\Cref{fig:remote}.
Specifically, let the source be modeled by a sequence of independent and identically distributed random variables $X.$
The source is first passed through a memoryless channel to form the observed sequence $Y.$ Observing only the sequence $Y,$ the encoder produces a compressed representation.
From the compressed representation, the original source sequence needs to be reconstructed to within a prescribed distortion level $\Delta.$ How many bits of coding rate are required to attain this goal?
The answer to this question is the remote rate-distortion function. A careful discussion of this can be found in~\cite[Section 3.5]{BergerBook}.
For the case where the distortion measure is the mean-squared error, the remote rate-distortion function is given by
\begin{align}
  R_X^R(\Delta) &= \inf_{p(\hat{x}|y): {\mathbb E}[d(X, \hat{X})] \le \Delta} I(Y; \hat{X}).
\end{align}
Berger, in~\cite[Section 3.5]{BergerBook}, shows a general technique to reduce the remote to a regular rate-distortion function,
albeit with a different distortion measure.
Indeed, the main vehicle is to introduce what Berger refers to as the {\it modified} distortion measure:
the original $d(x, \hat{x})$
is converted into a measure expressed
in terms of the remote source observation $y$ as
$d^* ( y, \hat{x} ) = {\mathbb E}_{X|y} \left[ d (X, \hat{x} ) \right]$.
This important insight shows that we might equivalently write the remote rate-distortion function as
\begin{align}
  R_X^R(\Delta) &= R_Y^*(\Delta) = \inf_{p(\hat{x}|y): {\mathbb E}[d^*(Y, \hat{X})] \le \Delta} I(Y; \hat{X}),
\end{align}
which we recognize to be a regular rate-distortion function for the source $Y,$ subject to a distortion criterion based on $d^* ( y, \hat{x} ),$ hence to asterisk in the superscript of $R_Y^*(\Delta).$
For the purpose of our consideration, we can further simplify (leveraging the well-known orthogonality property of conditional expectation)
\begin{align}
  d^*(y, \hat{x}) &= {\mathbb E}_{X|y} [ ( X - v(y))^2] +  ( v(y)- \hat{x})^2,
\end{align}
where we use the shorthand $v(y)={\mathbb E}[X|Y=y]$ for ease of notation. Therefore, we can rewrite
\begin{align}
  {\mathbb E}[d^*(Y, \hat{X})] &= \Delta_0 + {\mathbb E}[ (V(Y)- \hat{X})^2],
\end{align}
where $\Delta_0$ denotes the minimum mean-squared error as defined in~\Cref{eq:def:mmse}.
Observing that $\hat{X}$ is a dummy variable and leveraging the data processing inequality, Berger's elegant decomposition can be simplified to the following form:
\begin{align}
  R_X^R(\Delta) &= R_V(\Delta-\Delta_0) = \inf_{p(\hat{v}|v): {\mathbb E}[(V-\hat{V})^2] \le \Delta-\Delta_0} I(V; \hat{V}),\label{eqn-remote-SLB-almost}
\end{align}
which is the regular quadratic rate-distortion function for the random variable $V={\mathbb E}[X|Y].$
Therefore,~\Cref{thm-SLB-basic} now directly implies the following well known bounds.
\begin{Theorem}
The remote rate-distortion function subject to mean-squared error satisfies the following lower and upper bounds, using the shorthand $V={\mathbb E}[X|Y]:$
\begin{align}
 \frac{1}{2} \log^+ \frac{\entp{V}}{\Delta-\Delta_0} \le  R_X^R(\Delta)  \le \frac{1}{2} \log^+ \frac{\sigma_V^2}{\Delta-\Delta_0}, \label{eqn-remote-SLB}
\end{align}
for $\Delta > \Delta_0,$ where $\Delta_0 = {\mathbb E}\left[ \left(X - V\right)^2  \right].$
\end{Theorem}
As a side remark, we point out that a hand-tailored argument leading to~\Cref{eqn-remote-SLB-almost} for the special case of the mean-squared error was also given by Wolf and Ziv~\cite{WolfZ:70}, see e.g.~\cite[Eqn.(42)]{EswaranG:19}.

It is instructive to consider the special case where the noisy source observation is obtained by adding independent noise to the original source. That is, let $Y=X+Z,$ where $Z$ follows an arbitrary distribution with variance $\sigma_Z^2$ but is independent of $X.$ 
This case is well understood in the literature, see e.g.~\cite[Eqns.(17)-(18)]{EswaranG:19},
and is sometimes referred to as the additive noise remote source coding problem.
For this case, one can combine Inequalities~\eqref{eqn-remote-SLB} and~\eqref{eqn-estimation}
to obtain the following more explicit pair of bounds:

\begin{Theorem}\label{thm:SLB:remote:AN}
The additive noise remote rate-distortion function subject to mean-squared error satisfies the following lower bound, using the shorthand $V={\mathbb E}[X|Y]:$
 \begin{align}
R_X^R(\Delta) & \ge \frac{1}{2} \log^+ \frac{\entp{V}}{\Delta}  + \frac{1}{2} \log^+  \frac{ \entp{Y}}{\entp{Y}-\frac{\entp{X}}{\Delta}\entp{Z} } \label{eqn-remote-SLB-explicit-lower}
\end{align}
for $\Delta > \entp{X}\entp{Z}/\entp{Y},$ as well as the upper bound, 
 \begin{align}
 R_X^R(\Delta) &\le \frac{1}{2}\log^+\frac{\sigma_V^2}{\Delta}  + \frac{1}{2}\log^+ \frac{\sigma_Y^2}{\sigma_Y^2 - \frac{\sigma_X^2}{\Delta}{\sigma_{{Z}}^2}}, \label{eqn-remote-SLB-explicit-upper} \end{align}
for  $\Delta > \sigma_X^2\sigma_Z^2/\sigma_{Y}^2.$
 \end{Theorem}

There are conditions on the value of $\Delta$ in~\Cref{thm:SLB:remote:AN}.
To interpret them, we start by observing that only values $\Delta > \Delta_0$ are relevant, where $\Delta_0 = {\mathbb E}[ \left(X - V\right)^2].$ On account of~\Cref{eqn-estimation}, we can observe that the lower bound (\Cref{eqn-remote-SLB-explicit-lower}) is valid for all $\Delta$ of interest, but the upper bound (\Cref{eqn-remote-SLB-explicit-upper}) only holds for distortion values $\Delta$ larger than the {\it linear} minimum mean-squared error at which $X$ can be estimated from the noisy observation $Y.$ 
 
 An even more direct pair of bounds has been given for the additive noise remote source coding problem when the additive noise is Gaussian (and the original source distribution is arbitrary).
 This version is sometimes referred to as the {\it AWGN remote source coding problem.}
 For this case, a recent result enables a new proof. Namely, we can use~\cite[Theorem 1]{AtalikKG:22}, which says that if $Y=X+W,$ where $W$ is Gaussian noise, then we must have $h({\mathbb E}[X|Y]) +h(Y) \ge 2h(X).$
 This permits to take care of the term $\entp{V}$ in~\Cref{thm:SLB:remote:AN}.
 In this fashion, we obtain the following more direct pair of bounds.
  
\begin{Corollary}\label{thm:SLB:remote:AWGN}
The AWGN remote rate-distortion function subject to mean-squared error satisfies the following lower bound:
\begin{align}
R_X^R(\Delta)&\ge \frac{1}{2} \log^+ \frac{\entp{X}}{\Delta}  + \frac{1}{2} \log^+  \frac{ \entp{X}}{\entp{Y}-\frac{\entp{X}}{\Delta}\sigma_{Z}^2 }, \label{eq-EPI-explicit-lower}
\end{align}
for $\Delta > \entp{X}\sigma_Z^2/\entp{Y},$ as well as the upper bound
\begin{align}
  R_X^R(\Delta) &\le \frac{1}{2}\log^+\frac{\sigma_X^2}{\Delta}  + \frac{1}{2}\log^+ \frac{\sigma_X^2}{\sigma_Y^2 - \frac{\sigma_X^2}{\Delta}{\sigma_{{Z}}^2}}, \label{eq-EPI-explicit-upper}
\end{align}
for $\Delta > \sigma_X^2\sigma_Z^2/\sigma_{{Y}}^2.$
\end{Corollary}
There are conditions on the value of $\Delta$ in~\Cref{thm:SLB:remote:AWGN}.
These conditions can be interpreted exactly as in the discussion following~\Cref{thm:SLB:remote:AN}.
We also remark that this corollary can be established directly using a rather different proof technique via entropy power inequalities, see~\cite[Eqns.(48)-(52)]{EswaranG:19}.

\section{The Gray-Wyner Source Coding Network}

\begin{figure}[h!] 
\centering
\begin{tikzpicture}[scale=0.8]
\draw[black,thick] (0,1) rectangle (1,2);
\draw[black,thick,anchor=east] (-0.5,1.5) node{$(X_1,X_2)$};
\draw[black,thick,anchor=west] (5.5,3) node{$\hat{X}_1$};
\draw[black,thick,anchor=west] (5.5,0) node{$\hat{X}_2$};
\draw[black,thick,anchor=center] (0.5,1.5) node{$\mathcal{E}$};
\draw[black,thick,anchor=south] (2,0) node{$R_2$};
\draw[black,thick,anchor=south] (2,1.5) node{$R_c$};
\draw[black,thick,anchor=south] (2,3) node{$R_1$};
\draw[black,thick](-0.5,1.5)--(0,1.5);
\draw[->,black,thick](0.5,1)--(0.5,0)--(3.5,0);
\draw[->,black,thick](1,1.5)--(4,1.5)--(4,2.5);
\draw[->,black,thick](4,1.5)--(4,0.5);
\draw[->,black,thick](0.5,2)--(0.5,3)--(3.5,3);
\draw[black,thick,anchor=center] (4,0) node{$\mathcal{D}_2$};
\draw[black,thick,anchor=center] (4,3) node{$\mathcal{D}_1$};
\draw[black,thick] (3.5,-0.5) rectangle (4.5,0.5);
\draw[black,thick] (3.5,2.5) rectangle (4.5,3.5);
\draw[->,black,thick](4.5,0)--(5.5,0);
\draw[->,black,thick](4.5,3)--(5.5,3);
\end{tikzpicture}
\caption{The Gray-Wyner Network} 
\label{fig:Gray-Wyner}
\end{figure}
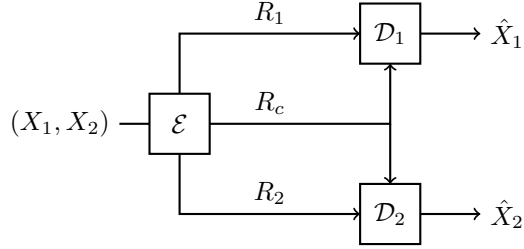

In this section, we encounter a source coding network scenario for which Shannon bounds can be given,
the so-called Gray-Wyner source coding network~\cite{Gray--Wyner74}, illustrated in~\Cref{fig:Gray-Wyner}.
This network was first proposed in~\cite{Gray--Wyner74}.
In this network, there is a single encoder, having access to two sources $X_1$ and $X_2,$ respectively.
The two sources are generally correlated with each other.
There are two decoders, and each decoder is interested in only one of the two sources.
The encoder produces three messages. One of these, of rate $R_c$ and termed the ``common'' message, is provided to both encoders.
The other two messages, of rates $R_1$ and $R_2,$ respectively, are provided to decoders 1 and 2, respectively.
The resulting region of attainable rate-distortion tuples is unknown except for special cases. These include the case where $(X_1,X_2)$ are jointly Gaussian and the distortion criterion is the mean-squared error, a case that was partially resolved in~\cite{Xu--Liu--Chen,ViswanathaAR:14} and fully determined in~\cite{Sula-Gastpar22}.

For ease of exposition, we make two simplifications in the sequel.
First, we focus on the total private rate delivered to the receivers. With reference to~\Cref{fig:Gray-Wyner}, we consider
\begin{align} 
R_1 + R_2 & \le R_p.
\end{align}
Second, we assume that the target distortion at the two receivers is the same and denote it by $\Delta.$
We study the smallest possible common rate $R_c$ that allows to attain a distortion of $\Delta$ at each decoder and requires a total private rate of at most $R_p.$
The advantage of these simplifications is that we can reduce the problem description to a compact form.
Namely, the resulting rate-distortion function can be written as
\begin{align} \label{eqn:GWdef}
R_c(\Delta, R_p) &= \inf_{W:R_{X_1|W}(\Delta)+R_{X_2|W}(\Delta)\leq R_p} I(X_1,X_2;W).
\end{align}
This rate-distortion function can be bounded in a way akin to the Shannon bounds.
In the present paper, we restrict attention to the case where the underlying source is symmetric (that is, where both source components have the same variance). More general cases will be treated elsewhere.
\begin{Theorem} \label{thm:main}
Let $(X_1,X_2)$ be an arbitrary source with correlation coefficient $\rho$ and where each component has variance $\sigma^2.$ Let the distortion measure be the mean-squared error. Then,
\begin{align} \label{eqn:partoneLUB}
&\frac{1}{2} \log^{+}{\frac{\entpsq{X_1,X_2}}{\sigma^2(1-|\rho|) \left( 2\Delta e^{R_p}+\sigma^2(|\rho|-1) \right)}} \leq R_c(\Delta, R_p)   \leq \frac{1}{2} \log^{+}{\frac{\sigma^2(1-\rho^2)}{(1-|\rho|) \left( 2\Delta e^{R_p}+\sigma^2(|\rho|-1) \right)}}
\end{align}
for $\sigma^2(1-|\rho|) \le \Delta e^{R_p} \le \sigma^2$ and
\begin{align} \label{eqn:parttwoLUB}
 \frac{1}{2} \log^{+}{\frac{\sigma^2\entpsq{X_1,X_2}}{\Delta^2 e^{2R_p}}} \leq R_c(\Delta, R_p) \leq \frac{1}{2} \log^{+}{\frac{\sigma^4(1-\rho^2)}{\Delta^2 e^{2R_p}}} 
\end{align}
for $\Delta e^{R_p} \le \sigma^2(1-|\rho|).$
\end{Theorem}
This result constitutes the novel technical contribution of this paper.
The proof of the lower bound in this theorem hinges on Brascamp-Lieb-style inequalities as given in~\Cref{lemma:thm:main} and~\Cref{lemma:thm:main:2} below.
A detailed proof is included in Appendices \ref{app:proofmainthmLB} and \ref{app:proofmainthmUB}.

To discuss this result, observe that 
 both in~\Cref{eqn:partoneLUB} and in~\Cref{eqn:parttwoLUB}, the difference of the lower and upper bounds is the Kullback-Leibler divergence between the distribution of $(X_1,X_2)$ and a bi-variate Gaussian distribution with the same covariance matrix as $(X_1,X_2).$
 On the one hand, this illustrates that~\Cref{thm:main} provides the exact characterization for the special case where $(X_1,X_2)$ are jointly Gaussian random variables. On the other hand, it shows that the bounds provided by~\Cref{thm:main} remain relevant for all distributions that are close (in Kullback-Leibler divergence) to a Gaussian distribution.

\section{The AWGN CEO Problem}

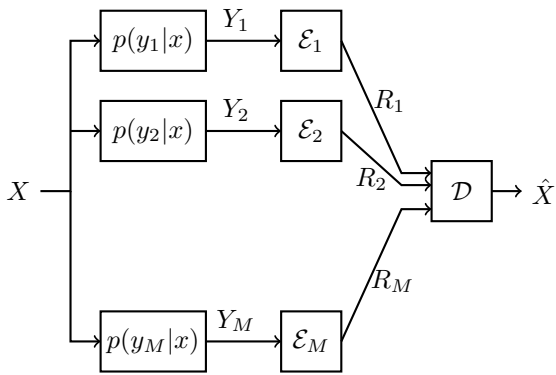
\begin{figure}[h!] 
\centering
\begin{tikzpicture}[scale=0.8]
\draw[black,thick,anchor=east] (-0.5,1.5) node{$X$};

\draw[->,black,thick](-0.5,1.5)--(0,1.5)--(0,4)--(0.5,4);
\draw[black,thick,anchor=center] (1.375,4) node{$p(y_1|x)$};
\draw[black,thick] (0.5,3.5) rectangle (2.25,4.5);
\draw[->,black,thick](2.25,4)--(3.5,4);
\draw[black,thick,anchor=south] (2.75,4) node{$Y_1$};
\draw[black,thick,anchor=center] (4,4) node{$\mathcal{E}_1$};
\draw[black,thick] (3.5,3.5) rectangle (4.5,4.5);

\draw[->,black,thick](0,2.5)--(0.5,2.5);
\draw[black,thick,anchor=center] (1.375,2.5) node{$p(y_2|x)$};
\draw[black,thick] (0.5,2) rectangle (2.25,3);
\draw[->,black,thick](2.25,2.5)--(3.5,2.5);
\draw[black,thick,anchor=south] (2.75,2.5) node{$Y_2$};
\draw[black,thick,anchor=center] (4,2.5) node{$\mathcal{E}_2$};
\draw[black,thick] (3.5,2) rectangle (4.5,3);

\draw[->,black,thick](0,1.5)--(0,-1)--(0.5,-1);
\draw[black,thick,anchor=center] (1.375,-1) node{$p(y_M|x)$};
\draw[black,thick] (0.5,-1.5) rectangle (2.25,-0.5);
\draw[->,black,thick](2.25,-1)--(3.5,-1);
\draw[black,thick,anchor=south] (2.75,-1) node{$Y_M$};
\draw[black,thick,anchor=center] (4,-1) node{$\mathcal{E}_M$};
\draw[black,thick] (3.5,-1.5) rectangle (4.5,-0.5);

\draw[->,black,thick](4.5,4)--(5.5,1.8)--(6,1.8);
\draw[black,thick,anchor=center] (5.3,3) node{$R_1$};
\draw[->,black,thick](4.5,2.5)--(5.5,1.6)--(6,1.6);
\draw[black,thick,anchor=center] (5,1.7) node{$R_2$};
\draw[->,black,thick](4.5,-1)--(5.5,1.2)--(6,1.2);
\draw[black,thick,anchor=center] (5.35,0) node{$R_M$};

\draw[black,thick] (6,1) rectangle (7,2);
\draw[black,thick,anchor=center] (6.5,1.5) node{$\mathcal{D}$};
\draw[->,black,thick](7,1.5)--(7.5,1.5);
\draw[black,thick,anchor=west] (7.5,1.5) node{$\hat{X}$};

\end{tikzpicture}
\caption{The $M$-agent CEO problem. $X$ is an arbitrary source with variance (power) $\sigma_X^2$ (not necessarily Gaussian) and entropy power $\entp{X}.$ For the version considered in the present paper, the source observation kernels $p(y_m|x)$ consist in adding independent Gaussian noises $Z_m$ of variance $\sigma_Z^2.$}\label{fig:CEO}
\end{figure}

The CEO source coding problem was introduced by Berger and collaborators in 1996~\cite{BergerZV:96}.
It is illustrated in Figure~\ref{fig:CEO}.
The CEO problem is one of several natural network extension of the remote source coding problem that we discussed in Section~\ref{Sec-remote} above.
In the CEO problem, multiple encoding devices each observe noisy (partial, distorted) versions of one and the same underlying source, characterized by the observation kernels $p(y_m|x),$ for $m=1, 2, \ldots, M,$
as illustrated in Figure~\ref{fig:CEO}.
In general, this is a formidable problem and has resisted a solution.
It is not possible to give a simple scalar optimization problem akin to the optimization problems that we considered in the preceding sections. 
For special cases, more progress has been made.
Most relevant to our discussion here is the work of Berger and Vishwanathan who made fundamental contributions towards resolving the problem in the special case of Gaussian statistics and quadratic distortion~\cite{ViswanathanB:97}.
Namely, with respect to Figure~\ref{fig:CEO}, in the quadratic Gaussian CEO problem, the underlying source $X$ is assumed to be Gaussian and all of the observation kernels $p(y_m|x)$ consist in adding (independent) Gaussian noises. Finally, the distortion criterion of interest is the mean-squared error between the underlying source $X$ and the reconstruction $\hat{X}.$
This version is the starting point for the result that we include next.
Specifically, the only change is that we allow the underlying source $X$ to have an arbitrary distribution, not necessarily Gaussian (but of finite variance $\sigma_X^2$ and entropy power $\entp{X}$).
We point to~\cite{EswaranG:19} (and the references therein) for a precise problem statements along with formal definitions.
Shannon-type bounds for this problem are established in~\cite{EswaranG:19}.
For the purpose of the present paper, we only cite the symmetric version since this is the most compact to state.

\begin{Theorem}[Corollary 2 in~\cite{EswaranG:19}]\label{thm-AWGN-CEO}
For the $M$-agent AWGN CEO problem with an arbitrary continuous underlying source $X,$ constrained to having finite differential entropy, 
with observation noise variance $\sigma_{Z_m}^2=\sigma_Z^2,$ for $m=1, 2, \ldots, M,$ and subject to mean-squared error distortion, the CEO sum-rate distortion function is lower bounded by
\begin{align}
R_X^{CEO}(\Delta) \geq R_{X, lower}^{CEO}(\Delta) & = \frac{1}{2} \log^+ \frac{\entp{X}}{\Delta}  + \frac{M}{2} \log^+ \frac{M \entp{X}}{M\entp{Y}-\frac{\entp{X}}{\Delta}\sigma_{Z}^2 } \label{eq:ceo_rd_lowerbd_sqer}
\end{align}
for $\Delta > \entp{X}\sigma_{Z}^2/(M\entp{Y}).$
Moreover, in this case, the CEO sum-rate distortion function is upper bounded by
\begin{align}
R_{X}^{CEO}(\Delta)  \leq R_{X,upper}^{CEO}(\Delta)
 & = \frac{1}{2}\log^+\frac{\sigma_X^2}{\Delta} + \frac{M}{2}\log^+ \frac{M\sigma_X^2}{M\sigma_{Y}^2 - \frac{\sigma_X^2}{\Delta}{\sigma_{{Z}}^2}} ,
\label{eq:ceo_rd_upperbd_sqer}
\end{align}
for $\Delta > \sigma_X^2\sigma_Z^2/(M\sigma_{Y}^2),$
where $ Y = \frac{1}{M} \sum_{m=1}^M Y_m = X + \frac{1}{M} \sum_{m=1}^M Z_m.$
\end{Theorem}
We note that like~\Cref{thm:main}, this theorem cannot be established with elementary techniques. Its proof involves generalizations of the entropy-power inequality. A detailed proof is given in~\cite{EswaranG:19}.

There are conditions on the value of $\Delta$ in~\Cref{thm-AWGN-CEO}.
To interpret them, we start by observing that only values $\Delta > \Var(X|Y_1, Y_2, \ldots, Y_M)$ are relevant. On account of~\Cref{eqn-estimation}, we can observe that the lower bound (\Cref{eq:ceo_rd_lowerbd_sqer}) is valid for all $\Delta$ of interest, but the upper bound (\Cref{eq:ceo_rd_upperbd_sqer}) only holds for $\Delta$ larger than the {\it linear} minimum mean-squared error at which $X$ can be estimated from the noisy observations $(Y_1, Y_2, \ldots, Y_M).$ 
While for the present survey, we only include this simplified statement for the sum-rate and for the symmetric case, we point out that a much more general result appears in~\cite{EswaranG:19}.

\section{Discussion and Open Problems}

The main purpose of this paper is to survey bounds of the Shannon type on quadratic rate-distortion problems and to put these bounds in the context of Berger's work on this topic. Specifically, the paper includes Shannon bounds for the conditional rate-distortion function, the Wyner-Ziv rate-distortion function, and the remote (or indirect) rate-distortion function. Proofs are given following Berger's techniques. Moreover, we also present a new pair of Shannon bounds for the Gray-Wyner network, and we include similar bounds for a special case of the CEO problem.

There are many open problems of interest.
In the present paper, we have focused on quadratic rate-distortion problems, {\it i.e.,} when the distortion measure is the mean-squared error.
For the standard direct rate-distortion problem, Berger~\cite[Section 4.3.1]{BergerBook} gives a much more general result, considering general difference distortion measures, that is, if the distortion measure is of the form $d(x,\hat{x}) = \rho(x-\hat{x})$ for some function $\rho(\cdot).$
Clearly, it would be of interest to extend the theorems surveyed in this paper beyond the quadratic case.
Another avenue of research comes out of Berger's variational formula for the rate-distortion function. In the present paper, we have seen how it can be leveraged directly in cases including the conditional rate-distortion function or the remote rate-distortion function. An extension to source coding networks would be of great interest.
For the networks discussed in this survey, an important avenue for future research would be to extend the bounds for the AWGN CEO problem to cases where the observation noise is not Gaussian.
Finally, bounds akin to the ones discussed here ought to be available for the standard distributed lossy source coding problem spearheaded by Berger and Tung~\cite{berger-msc_lecturenotes,tungthesis78}.

\appendices

\section{Berger's variational theorem}\label{app-BergerProofSLB}

To establish the Shannon lower bound,
Berger uses a proof technique that lends itself to further generalization.
Specifically, Berger establishes the following alternative form of the rate-distortion function $R_X(\Delta)$ defined in~\Cref{Eq-basicRD}:

\begin{Theorem}[Theorem 4.2.3 in~\cite{BergerBook}]\label{thm:Bergervariational}
Let $\Lambda_s$ be the set of all nonnegative functions $\lambda(x)$ satisfying
\begin{align}
  \int  \lambda(x) p(x) e^{sd(x,y)} dx & \le 1,
\end{align}
for all values of $y.$ Then,
\begin{align}
 R_X(\Delta) & = \sup_{s \le 0, \lambda(x) \in {\Lambda}_s} \left( s\Delta + \int  p(x) \log \lambda(x)dx \right).
\end{align}
\end{Theorem}

Using this theorem, it is possible to give lower bounds to the rate-distortion function simply by selecting a valid function $\lambda(x)$ and a non-positive real number $s.$
Specifically, we may choose (see~\cite[Sec.4.3.1]{BergerBook}
\begin{align}
  \lambda(x) &= \frac{K}{p(x)}.
\end{align}
Then, for quadratic distortion $d(x,y)=(x-y)^2,$ we select $K=\sqrt{\frac{-s}{\pi}},$
which can be verified to satisfy the constraint of the theorem for all negative numbers $s.$
Moreover, we select $s=-\frac{1}{2\Delta}.$
Plugging in directly leads to the lower bound formula given in~\Cref{thm-SLB-basic}.

\section{Proofs of~\Cref{lem:cond-rate-distor} and~\Cref{lem:cond-rate-distor-WZ}}\label{app:proof:lem:cond-rate-distor}

To establish the lower bound, we offer a proof based on Berger's variational theorem as discussed in Appendix~\ref{app-BergerProofSLB}.
To leverage this theorem, we start by observing that we can write
\begin{align}
    R_{X|W}(\Delta) &= \min_{\Delta_W: {\mathbb E}_W[\Delta_W]\le \Delta} {\mathbb E}_W[R_{X|W}(\Delta_W)].
\end{align}
For a fixed realization $W=w,$ the expression inside the last expectation is merely a standard rate-distortion function (concerning the conditional distribution $p(x|w)$).
Hence, using~\Cref{thm:Bergervariational}, we can write
\begin{align}
   R_{X|W}(\Delta) &= \min_{\Delta_W: {\mathbb E}_W[\Delta_W]\le \Delta}
     {\mathbb E}_W\left[\sup_{\substack{s_W \le 0,\\ \lambda_W(x) \in {\Lambda}_{s_W}(W)}} \left( s_W\Delta_W + \int  p(x|W) \log \lambda_W(x) dx \right) \right],
\end{align}
where $\Lambda_{s}(w)$ is the set of all nonnegative functions $\lambda_w(x)$ satisfying
\begin{align}
  \int  \lambda_w(x) p(x|w) e^{sd(x,y)} dx & \le 1,
\end{align}
for all values of $y.$
Now, selecting $s_w \equiv s,$ we get the lower bound
\begin{align}
   R_{X|W}(\Delta) & \ge \sup_{s \le 0} \min_{\Delta_W: {\mathbb E}_W[\Delta_W]\le \Delta} {\mathbb E}_W\left[  s\Delta_W +  \sup_{\lambda_W(x) \in {\Lambda}_{s}(W)} \int  p(x|W) \log \lambda_W(x) dx \right] \\
    &= \sup_{s \le 0}\min_{\Delta_W: {\mathbb E}_W[\Delta_W]\le \Delta} \Biggl\{  s{\mathbb E}_W[\Delta_W] + {\mathbb E}_W\left[\sup_{\lambda_W(x) \in {\Lambda}_{s}(W)}\int  p(x|W) \log \lambda_W(x) dx \right]  \Biggr\} \\
    &  \ge \sup_{s \le 0} s\Delta + {\mathbb E}_W\left[\sup_{\lambda_W(x) \in {\Lambda}_{s}(W)}\int  p(x|W) \log \lambda_W(x) dx \right]  .
\end{align}
To obtain a lower bound, it now suffices to choose $s$ and $\lambda_w(x)$ wisely.
Following the proceedings in Appendix~\ref{app-BergerProofSLB}, we select
\begin{align}
    \lambda_w(x) &=  \frac{K}{p(x|w)},
\end{align}
and we select $K=\sqrt{\frac{-s}{\pi}}.$
For this choice, recalling that we consider quadratic distortion $d(x,y)=(x-y)^2,$ it is straightforward to verify that $\lambda_w(x)\in \Lambda_{s}(w)$ for every $w.$
Finally, we select $s=-\frac{1}{2\Delta}$ to obtain the claimed lower bound.

We also note that an alternative way of proving the lower bound leverages~\cite[Theorem 3.1]{Gray73} which establishes that $R_{X|W}(\Delta) \ge R_X(\Delta) - I(X;W).$ Combining this with~\Cref{thm-SLB-basic} leads to the claimed lower bound, see e.g.~\cite[App. A]{SulaG:22isit}.

For the upper bound, we start by leveraging the following simple facts (established for example in~\cite{Gray--1972}):
\begin{align}
 R_{X|W}(\Delta) &= R_{X-{\mathbb E}[X|W]|W}(\Delta) \\
    & \le R_{X-{\mathbb E}[X|W]}(\Delta)
\end{align}
At this point, we can again leverage~\Cref{thm-SLB-basic}. The only remaining argument is to find the variance of the random variable $X-{\mathbb E}[X|W].$
Since the mean of this random variable vanishes, the variance is simply
 ${\mathbb E}[ \left( X-{\mathbb E}[X|W] \right)^2 ],$
which, in the notation used in the present paper, we choose to denote as $\Var(X|W).$
Hence,
\begin{align}
 R_{X|W}(\Delta)    & \le  \frac{1}{2}\log^+{\frac{\Var(X|W)}{\Delta}},
\end{align}
which completes the proof for the conditional rate-distortion function.

For the upper bound in~\Cref{lem:cond-rate-distor-WZ}, it suffices to plug in a judicious choice of auxiliary $U$
and upper bound the resulting mutual information and distortion.
Recall that $U$ must be chosen to satisfy the Markov constraint $W-X-U.$
Hence, let us choose $U=\tilde{\rho} X + \sqrt{1-\tilde{\rho}^2}Z,$ where $Z$ is a zero-mean Gaussian of variance $\sigma_X^2$ independent of $X$ and $W.$ Note that with this choice, the correlation coefficient between $X$ and $U$ is precisely given by $\tilde{\rho}.$
For this choice of the auxiliary, we can upper bound the incurred distortion by analyzing
the optimal {\it linear} estimator $\hat{X}_{lin}(U,W)$ of $X$ given both $U$ and $W.$ 
This distortion is well known and can be expressed as
\begin{align}
\frac{\mathbb{E}[(X-\hat{X})^2]}{\sigma_X^2} &\le \frac{\mathbb{E}[(X-\hat{X}_{lin}(U,W))^2]}{\sigma_X^2} =  \frac{1}{1+\gamma+\tilde{\gamma}},
\end{align}
where $\gamma=\rho^2/(1-\rho^2)$ and $\tilde{\gamma}=\tilde{\rho}^2/(1-\tilde{\rho}^2).$ This can be seen for example by maximum ratio combining.
Setting this distortion expression equal to $\Delta/\sigma_X^2$ gives the condition $\tilde{\gamma}=\frac{\sigma_X^2}{\Delta} -1 - \gamma.$
Moreover, for this $U,$ we observe by standard arguments that
\begin{align}
 I(X;U|W) &= h(U|W) - h(U|X).
\end{align}
By our construction, the second term is merely the entropy of a Gaussian random variable.
For the first term,
from the conditional maximum entropy theorem (see e.g.~\cite[Lemma 1]{Thomas:87}),
we have $h(U|W) \le h(U^*|W^*),$ where $(U, W)$ and $(U^*, W^*)$ have the same covariance matrix,
but $(U^*, W^*)$ are jointly Gaussian. For jointly Gaussian random variables, entropy formulas are of course well known and allow us to write
\begin{align}
 I(X;U|W) & \le \frac{1}{2} \log \left( 1 + (1-\rho^2) \tilde{\gamma} \right).
\end{align}
Combining gives the claimed upper bound.

\section{Proof of~\Cref{thm:main}, Part I : Lower Bound} \label{app:proofmainthmLB}

This result constitutes a novel contribution of the present paper, having previously appeared in the conference version only~\cite{SulaG:22isit}.
Therefore, a detailed proof is included.
Without loss of generality, in the proof, we assume that $0\le \rho \le 1$ (that is, non-negative correlation) to avoid notational clutter.
Let us start from $R_c(R_p)$ that we defined in (\ref{eqn:GWdef}),
\begin{align}
R_c( R_p)&=\inf_{W:R_{X_1|W}(\Delta)+R_{X_2|W}(\Delta)\leq R_p} I(X_1,X_2;W) .
\end{align}
The first step is to relax the constraint.
Using the lower bound given in~\Cref{lem:cond-rate-distor},
we know that any $W$ for which we have $R_{X_1|W}(\Delta)+R_{X_2|W}(\Delta)\leq R_p$
must also satisfy $\frac{1}{2} \log^+{\frac{\entp{X_1|W}}{\Delta}}+ \frac{1}{2} \log^+{\frac{\entp{X_2|W}}{\Delta}} \leq R_p.$
Therefore, we can infer that
\begin{align}
R_c( R_p) & \geq \inf_{W:\frac{1}{2} \log^+{\frac{\entp{X_1|W}}{\Delta}}+ \frac{1}{2} \log^+{\frac{\entp{X_2|W}}{\Delta}} \leq R_p} I(X_1,X_2;W). \label{eqn:defcondratedist} 
\end{align}
From weak duality, this can be further lower bounded by
\begin{align}
R_c( R_p) & \geq \inf_W I(X_1,X_2;W) + \frac{\nu}{2} \log^+{\frac{\entp{X_1|W}}{\Delta}} \nonumber \\
& \quad \quad + \nu \left( \frac{1}{2} \log^+{\frac{\entp{X_2|W}}{\Delta}} - R_p \right). \label{eqn:weakduality}
\end{align}
To continue, let us now consider the case where the distortion satisfies $\Delta \le \min \{ \entp{X_1|W}, \entp{X_2|W} \},$
noting that the other cases are simple to handle and much less interesting.
In this case, the two $\log^+(\cdot)$ in the previous expression can be replaced by $\log(\cdot)$ and combined to yield the following quite canonical form:
\begin{align}
\lefteqn{R_c( R_p)  \geq h(X_1,X_2) - \nu R_p - \nu \log(2 \pi e \Delta)} \nonumber \\
&  + \nu \cdot  \inf_W \left\{ h(X_1|W)+h(X_2|W)- \frac{1}{\nu}h(X_1,X_2|W)\right\},
\label{eqn:minsplit} 
\end{align}
where $\nu$ is the Lagrangian parameter that must satisfy $\nu\geq 0$. At this stage, we prefer to rewrite this in terms of normalized random variables $\tilde{X}_1=X_1/\sqrt{\sigma^2}$ and $\tilde{X}_2=X_2/\sqrt{\sigma^2},$ leading to
\begin{align}
\lefteqn{R_c( R_p)  \geq h(\tilde{X}_1,\tilde{X}_2) - \nu R_p - \nu \log(2 \pi e \Delta) + \nu \log \sigma^2 } \nonumber \\
& + \nu \cdot \inf_W  \left\{h(\tilde{X}_1|W)+h(\tilde{X}_2|W)- \frac{1}{\nu}h(\tilde{X}_1,\tilde{X}_2|W)\right\}. \label{eqn:minsplit2} 
\end{align}
Observe that the covariance matrix of the normalized random variables $(\tilde{X}_1, \tilde{X}_2)$ is given by
\begin{align}
    \begin{pmatrix} 1 & \rho \\ \rho &1  \end{pmatrix}.
\end{align}
To continue from here, we use the following fact, which is~\cite[Theorem~8]{Sula-Gastpar22}:

\begin{Lemma}\label{lemma:thm:main}
Let $K \succeq 0$ and let $(X, Y)\sim p_{X,Y}$ with covariance matrix $K.$
For all $0 < \lambda < 1$, there exists a $0\preceq K^{\prime} \preceq K$ and $(X^{\prime},Y^{\prime})\sim \mathcal{N}(0,K^{\prime})$ such that
\begin{align}
&\inf_W \left\{ h(X|W)+h(Y|W) - (1+\lambda) h(X, Y|W) \right\} \nonumber \\
& \quad \quad \geq h(X^{\prime})+ h(Y^{\prime}) -(1+\lambda)h(X^{\prime}, Y^{\prime}) . \label{Eq-Thm:Hypercontract}
\end{align}
\end{Lemma}

With this lemma, we thus have, for $\frac{1}{2} \le \nu \le 1,$
\begin{align}
R_c( R_p)& \geq h(\tilde{X}_1,\tilde{X}_2) - \nu R_p -\nu \log{(2 \pi e \frac{\Delta}{\sigma^2})}  \nonumber\\
& \quad \quad + \nu \cdot \hspace{-2em} \min_{0 \preceq K^{\prime} \preceq \begin{pmatrix} 1 & \rho \\ \rho &1  \end{pmatrix}} h(X^{\prime})+h(Y^{\prime})- \frac{1}{\nu}h(X^{\prime},Y^{\prime}) \label{eqn:twoboundeval} 
\end{align}
This minimum can be further lower bounded using~\cite[Lemma~13]{Sula-Gastpar22}, which says:
\begin{Lemma}\label{lemma:thm:main:2}
For $\lambda \leq \rho$, the following inequality holds
\begin{align} \label{eqn:covoptim}
&\min_{K^{\prime}: 0 \preceq K^{\prime} \preceq \begin{pmatrix} 1 & \rho \\ \rho &1  \end{pmatrix}} h(X^{\prime})+h(Y^{\prime})-(1+\lambda)h(X^{\prime},Y^{\prime}) \nonumber \\
&\hspace{4em} \geq  \frac{1}{2} \log{\frac{1}{1-\lambda^2}}-\frac{\lambda}{2} \log{(2\pi e)^2\frac{(1-\rho)^2(1+\lambda)}{1-\lambda}},
\end{align}
where $(X^{\prime},Y^{\prime})\sim \mathcal{N}(0,K^{\prime}).$
\end{Lemma}

Hence, we find for $\frac{1}{\nu}-1 \le \rho,$
\begin{align}
R_c( R_p)& \geq \frac{1}{2} \log{(2\pi e)^2\entpsq{\tilde{X}_1,\tilde{X}_2}} - \nu R_p -\nu \log{(2 \pi e \frac{\Delta}{\sigma^2})} \nonumber \\
& \quad \quad + \frac{\nu}{2} \log{\frac{\nu^2}{2\nu-1}}  -\frac{1-\nu}{2} \log{(2\pi e)^2\frac{(1-\rho)^2}{2\nu-1}}. \label{eqn:hypereval2}
\end{align}
The final step of the proof is to maximize this expression over all choices of $\nu$ in the interval for which our lower bound holds, that is,
in the interval $1 \geq \nu \geq \frac{1}{1+\rho}.$
To do so, consider the function
\begin{align}
\ell(\nu)&:=\frac{1}{2} \log{(2\pi e)^2\entpsq{\tilde{X}_1,\tilde{X}_2}} - \nu R_p -\nu \log{(2 \pi e \frac{\Delta}{\sigma^2})} \nonumber \\
&\quad \quad  + \frac{\nu}{2} \log{\frac{\nu^2}{2\nu-1}}-\frac{1-\nu}{2} \log{(2\pi e)^2\frac{(1-\rho)^2}{2\nu-1}},
\end{align}
for $1 \geq \nu \geq \frac{1}{1+\rho}.$
To goal is to maximize the function, i.e., to solve
$\max_{1 \geq \nu \geq \frac{1}{1+\rho}} \ell(\nu)$.
For notational compactness, let us define $\tilde{\Delta} = \frac{\Delta}{\sigma^2}.$
Note that the function $\ell$ is concave since
\begin{align}
\frac{\partial^2 \ell}{\partial \nu^2}=-\frac{1}{\nu(2\nu-1)}<0,
\end{align}
and by studying the monotonicity
\begin{align}
\frac{\partial \ell}{\partial \nu}=\log{\frac{\nu(1-\rho)}{(2\nu-1)\tilde{\Delta} e^{R_p}}},
\end{align}
its maximal value occurs when the derivative vanishes, that is, when $\nu_*=\frac{\tilde{\Delta} e^{R_p}}{2 \tilde{\Delta} e^{R_p}-1+\rho}.$ Substituting for the optimal $\nu_*$ we get
\begin{align}
R_c( R_p) &\geq \ell \left(\frac{\tilde{\Delta} e^{R_p}}{2\tilde{\Delta} e^{R_p}-1+\rho} \right) \\
&=\frac{1}{2} \log^+ \frac{\entpsq{\tilde{X}_1,\tilde{X}_2}}{(1-\rho) \left( 2 \tilde{\Delta} e^{R_p}-1+\rho \right) },
\end{align}
for $1 \geq \nu_* \geq \frac{1}{1+\rho}$, which means the bound is valid for the range $1-\rho \leq \tilde{\Delta} e^{R_p} \leq 1$.

We now separately also consider the case where $ \tilde{\Delta} e^{R_p} \leq 1-\rho$. In this case note that $\nu(1-\rho) \geq \nu \tilde{\Delta} e^{R_p} \geq (2\nu-1)\tilde{\Delta} e^{R_p}$ for $\nu \leq 1$. This implies $\frac{\nu(1-\rho)}{(2\nu-1) \tilde{\Delta} e^{R_p}} \geq 1$, thus we have $\frac{\partial \ell}{\partial \nu} \geq 0$. Since the function is concave and increasing the maximum is attained at $\nu_*=1$, thus the maximum is attained at the boundary, and we obtain the lower bound
\begin{align}
R_c(R_p) \geq \ell \left(1 \right) =\frac{1}{2} \log^+ \frac{\entpsq{\tilde{X}_1,\tilde{X}_2}}{\tilde{\Delta}^2 e^{2R_p}},
\end{align}
and this bound is valid for $\tilde{\Delta} e^{R_p} \leq 1-\rho$.
Combining, we thus get
\begin{align}
\lefteqn{R_c( R_p)=} \nonumber \\
&\left\{ \begin{array}{lr} \frac{1}{2} \log^{+}{\frac{\entpsq{\tilde{X}_1,\tilde{X}_2}}{(1-\rho) \left( 2\tilde{\Delta} e^{R_p}+\rho-1 \right)}}, &   \mbox{ if } 1-\rho \le \tilde{\Delta} e^{R_p} \le 1  \\
 \frac{1}{2} \log^{+}{\frac{\entpsq{\tilde{X}_1,\tilde{X}_2}}{\tilde{\Delta}^2 e^{2R_p}}} , &   \mbox{ if }  \tilde{\Delta} e^{R_p} \le 1-\rho.
\end{array} \right. \label{eqn:lastGrayWyner}
\end{align}
The claimed lower bound is now obtained by recalling that $\tilde{\Delta} = \Delta/\sigma^2$ and that $\entpsq{\tilde{X}_1,\tilde{X}_2} = \entpsq{X_1,X_2}/\sigma^2.$

\section{Proof of Theorem \ref{thm:main}, Part II : Upper Bound} \label{app:proofmainthmUB}

For the upper bound, as is typical for the Shannon bounds studied in this paper, the technique consists in selecting auxiliaries according to (conditional) Gaussian distributions.
For the Gray-Wyner network, two different cases need to be considered: the auxiliary could be a scalar (one-dimensional) random variable, or it could be a two-dimensional random vector.
We separately optimize each case, and then select the better of the two depending on the target distortion $\Delta.$ Intuitively, when the target distortion is large, then a one-dimensional auxiliary is better, and when the distortion is small, we need to consider two-dimensional auxiliaries. In the sequel, the optimizations are carried out in detail.
To keep notation simple, we write the proof assuming that the variances of $X$ and $Y$ are equal to 1.
Clearly, if a distortion $\Delta$ is attainable for sources of variance 1, then a distortion of $\Delta \sigma^2$ is attainable for sources of variance $\sigma^2.$

\subsection{$W$ is a scalar random variable}
We managed to show in Theorem \ref{lem:cond-rate-distor} that $R_{X_1|W}(\Delta) \leq \frac{1}{2} \log{\frac{\mathbb{E}[\Var(X_1|W)]}{\Delta}}$ and in order to satisfy the constraint in (\ref{eqn:GWdef}) we need to finally show that $\mathbb{E}[\Var(X_1|W)] \leq \Delta e^{R_p}$. We construct $W$ as
\begin{align}
W=\alpha(X_1+X_2) + N,
\end{align}
where $N$ is independent of $(X_1,X_2)$ and $N\sim \mathcal{N}\left(0,\frac{2\Delta e^{R_p}+\rho -1}{1+\rho}\right)$ and we choose $\alpha=\frac{\sqrt{1-\Delta e^{R_p}}}{1+\rho}$, thus
\begin{align}
\Var(X_1|W)&= \Var(X_1)-\Var(\mathbb{E}[X_1|W]) \\
&= \mathbb{E}[X_1^2] - \mathbb{E}[\mathbb{E}^2[X_1|W]] \\
& \leq \mathbb{E}[X_1^2] - \frac{\mathbb{E}^2[W\mathbb{E}[X_1|W]]}{\mathbb{E}[W^2]} \label{eqn:appchauchy} \\
&=\frac{\mathbb{E}[X_1^2]\mathbb{E}[W^2]-\mathbb{E}^2[X_1W]}{\mathbb{E}[W^2]}\\
&=\Delta e^{R_p} \label{eqn:subalphasqrt}
\end{align}
where \Cref{eqn:appchauchy} follows from the Cauchy–Schwarz inequality and \Cref{eqn:subalphasqrt} follows from $\mathbb{E}[X_1W]=\sqrt{1-\Delta e^{R_p}}$.
The same arguments can be applied to bound $\mathbb{E}[\Var(X_2|W)],$ and thus, $R_{X_2|W}(\Delta).$
Thus,
\begin{align}
R_c(R_p)\leq I(X_1,X_2;W)&=h(W)-h(N) \\
&\leq \frac{1}{2}\log{\frac{\Var(W)}{\Var(N)}} \label{eqn:thm1proofaux1} \\
&=\frac{1}{2}\log{\frac{1+\rho}{2\Delta e^{R_p}+\rho -1}} \label{eqn:thm1proofaux2},
\end{align}
where \Cref{eqn:thm1proofaux1} follows from $h(W) \leq \frac{1}{2} \log{(2\pi e \Var(W))}$ and \Cref{eqn:thm1proofaux2} follows from $\Var(W)=1$ and $\Var(N)=\frac{2\Delta e^{R_p}+\rho -1}{1+\rho}$. Note that \Cref{eqn:thm1proofaux2} corresponds to the upper bound in \Cref{eqn:partoneLUB}.

\subsection{$W$ is a two-dimensional random vector}

We construct $W=(W_1,W_2)$ as follows:
\begin{align}
 W_1 &= \alpha X_1 + \beta X_2 + N_1,  \\
 W_2 &= \beta X_1 + \alpha X_2 + N_2, \label{eqn:alphabeta}
\end{align}
where $N_1$ and $N_2$ are independent zero-mean Gaussians of variance $\Delta e^{R_p},$
independent of $(X_1,X_2).$
Moreover, the constants $\alpha$ and $\beta$ are selected to satisfy
\begin{align}
\alpha^2 + \beta^2 &= 1 - \frac{\Delta e^{R_p}}{1-\rho^2}, \label{eqn:ab1} \\
2\alpha\beta &= \rho \frac{\Delta e^{R_p}}{1-\rho^2}. \label{eqn:ab2}
\end{align}
With these choices, we have ${\mathbb E}[W_1^2] =  {\mathbb E}[W_2^2]=1$ and ${\mathbb E}[W_1W_2]=\rho$.
This permits to upper bound the minimum mean-squared error by the mean-squared error of the optimal {\it linear} estimator as
\begin{align}
  \Var(X_1|W_1,W_2) & \le \Delta e^{R_p}.
\end{align}
To see this, we start by observing that $\Var(X_1|W_1,W_2)$ is upper bounded by the distortion incurred by the optimal {\it linear} estimator, which is well known to be
\begin{align}
\Var(X_1|W_1,W_2) & \le
  \mathbb{E}[X_1^2]-
\begin{bmatrix}
\mathbb{E}[X_1W_1] \\
\mathbb{E}[X_1W_2] \\
\end{bmatrix}^T
\begin{bmatrix}
\mathbb{E}[W_1^2] & \mathbb{E}[W_1W_2]  \\
\mathbb{E}[W_1W_2] & \mathbb{E}[W_2^2] \\
\end{bmatrix}^{-1}
\begin{bmatrix}
\mathbb{E}[X_1W_1] \\
\mathbb{E}[X_1W_2] \\
\end{bmatrix}
\end{align}
To complete the proof, we can explicitly calculate:
\begin{align}
\mathbb{E}[X_1^2]-&
\begin{bmatrix}
\mathbb{E}[X_1W_1] \\
\mathbb{E}[X_1W_2] \\
\end{bmatrix}^T
\begin{bmatrix}
\mathbb{E}[W_1^2] & \mathbb{E}[W_1W_2]  \\
\mathbb{E}[W_1W_2] & \mathbb{E}[W_2^2] \\
\end{bmatrix}^{-1}
\begin{bmatrix}
\mathbb{E}[X_1W_1] \\
\mathbb{E}[X_1W_2] \\
\end{bmatrix} \\
&= 1-
\begin{bmatrix}
\alpha +\beta \rho \\
\beta +\alpha \rho \\
\end{bmatrix}^T
\begin{bmatrix}
1 & \rho  \\
\rho & 1 \\
\end{bmatrix}^{-1}
\begin{bmatrix}
\alpha +\beta \rho \\
\beta +\alpha \rho \\
\end{bmatrix} \\
&= 1-(\alpha^2+\beta^2) -2\alpha\beta\rho \\
&= \Delta e^{R_p}
\end{align}
by using (\ref{eqn:ab1}) and (\ref{eqn:ab2}). 
The validity region is $\Delta e^{R_p}< 1-\rho$, reflected in the upper bound in (\ref{eqn:parttwoLUB}).
Therefore, we can be sure that the private rate satisfies
\begin{align}
  R_{X_1|W_1,W_2}(\Delta) & \le \frac{1}{2} \log \frac{ \Var(X_1|W_1,W_2)}{\Delta} \le \frac{R_p}{2},
\end{align}
as desired.
The corresponding common rate $R_c$ can be bounded as follows:
\begin{align}
  R_c & \le I(X_1,X_2; W_1, W_2) \\
    & = h(W_1, W_2) -h(N_1, N_2) \\
    & \le \frac{1}{2} \log \frac{1-\rho^2}{\Delta^2 e^{2 R_p}}.
\end{align}
This completes the second half of the claimed upper bound.


\end{document}